\documentclass[a4paper,11pt]{article}
\usepackage{pos}
\usepackage{bm}

\title{High-energy QCD phenomenology with next-to-leading order
BFKL eigenfunctions}

\author*[a,b]{Ada Polizzi}
\author[c]{Michael Fucilla}
\author[a,b]{Alessandro Papa}

\affiliation[a]{Dipartimento di Fisica, Università della Calabria, \\ Arcavacata di Rende, 87036 Cosenza, Italy}

\affiliation[b]{INFN, Gruppo Collegato di Cosenza, \\ Arcavacata di Rende, 87036 Cosenza, Italy}

\affiliation[c]{Theoretical Physics Division, National Centre for Nuclear Research, Pasteura 7, Warsaw, 02-093, Poland}

\emailAdd{ada.polizzi@unical.it}
\emailAdd{michael.fucilla@unical.it}
\emailAdd{alessandro.papa@fis.unical.it}

\abstract{We discuss the phenomenological use of the next-to-leading order (NLO)
eigenfunctions of the Balitsky--Fadin--Kuraev--Lipatov (BFKL) kernel.
We first recall the BFKL representation of high-energy scattering
amplitudes and the role played by the eigenfunctions in the spectral
representation of the Green's function. We then compare two equivalent
representations of the next-to-leading logarithmic approximation (NLLA)
amplitude, respectively based on the leading-order (LO) and NLO eigenfunctions. The two formulations are equivalent
within NLLA, while differing by terms beyond this accuracy which could influence the phenomenological stability of
high-energy observables. The aim is to test the possible role of NLO-eigenfunctions-induced terms in improving this stability, discussing the amplitude dependence on the renormalization and
energy scales. As an application, we consider first the forward electroproduction of two light vector mesons and then we comment on preliminary applications to
Mueller--Navelet jet production.}

\FullConference{
The 33rd International Workshop on Deep Inelastic Scattering and Related Subjects (DIS2026)\\
4 - 8 May 2026\\
Bologna, Italy\\
}

\begin{document}
\maketitle

\section{Theoretical background}

\subsection{Introduction to BFKL}

At large hard scales and moderate values of
the longitudinal momentum fraction $x$, perturbative contributions are
dominated by collinear logarithms, which are resummed by the
Dokshitzer--Gribov--Lipatov--Altarelli--Parisi (DGLAP) evolution
equations. A different situation arises when the center-of-mass energy
becomes much larger than the hard scale. Since in deep inelastic
scattering $x$ decreases when the energy is increased at fixed
virtuality, high-energy collisions provide access to the small-$x$
region.

In the so-called semi-hard regime,
\begin{equation}
s \gg Q^2 \gg \Lambda_{\rm QCD}^2 ,
\end{equation}
the amplitude is governed by 
real gluons emissions strongly ordered in rapidity, while their transverse
momenta remain of comparable size, defining the Multi--Regge--Kinemtatics (MRK). Each large rapidity interval gives
rise to a logarithm of the energy, therefore the perturbative series is spoiled and large energy logarithms are resummed in the BFKL formalism
 \cite{BFKL1,BFKL2,BFKL3,BFKL4}.
The BFKL approach lays its foundation in Regge theory, which describes the high-energy behavior of scattering amplitudes in the Regge limit, where they
are dominated by the exchange of Reggeon in the $t$-channel.
For a generic Reggeon with trajectory $\alpha(t)$, the corresponding
high-energy behavior of the amplitude is schematically of the form
\begin{equation}
\mathcal A(s,t)\sim s^{\alpha(t)}.
\end{equation}
The exchange carrying vacuum quantum numbers is of particular
importance, since it gives the dominant contribution to total
cross sections at high energies. This exchange is traditionally
associated with the Pomeron.

In perturbative QCD, gluons are known to Reggeize and Reggeized gluons represent the elementary building blocks of the high-energy
description. In fact, in the color-singlet channel the perturbative realization
of the Pomeron is obtained from the exchange of two Reggeized gluons in
the $t$-channel. Their contrubution, together with the real
gluon emissions in the MRK, gives rise to a gluon ladder which describes the BFKL dynamics.

In this high-energy formalism it is possible to show that the scattering amplitude factorizes in transverse-momentum space and can be written as a convolution between two impact factors and a Green's function: 
\begin{equation}
{\rm Im}\mathcal A =
\frac{s}{(2\pi)^2}
\int\frac{d^2\bm q_1}{\bm q_1^2}
\int\frac{d^2\bm q_2}{\bm q_2^2}
\Phi_A(\bm q_1;s_0)
\int_{\delta-i\infty}^{\delta+i\infty}
\frac{d\omega}{2\pi i}
\left(\frac{s}{s_0}\right)^\omega
G_\omega(\bm q_1,\bm q_2)
\Phi_B(-\bm q_2;s_0).
\end{equation}
Impact factors $\Phi_A$ and $\Phi_B$ are process dependent: they
describe the transition between initial and final states including the coupling to the two-Reggeized-gluon exchange.
Conversely, the Green's function $G_\omega$ is universal and describes
the evolution in rapidity of the two-Reggeized-gluon system and of the whole scattering amplitude. The Green's function in fact resums contributions associated with
successive emissions along the BFKL gluon ladder, at any order in perturbation theory, and therefore reorganizes the $s$-dependence in terms proportional to $\alpha_s (\text{ln}s)$  at LLA and to $\alpha_s(\alpha_s (\text{ln}s))$ at NLLA.

This operation is realized by the BFKL equation:
\begin{equation}
\omega G_{\omega}(\bm q_1,\bm q_2)=\delta^{(2)}(\bm q_1-\bm q_2)
+
\int d^2\bm q_r,
K(\bm q_1,\bm q_r)
G_\omega(\bm q_r,\bm q_2),
\end{equation}
where the BFKL kernel represents the building block of high-energy
evolution and contains both virtual and real contributions. The virtual
part is related to Reggeization of the exchanged gluons, whereas the
real part describes real emissions of the ladder.

Writing the BFKL equations in the operator form
\begin{equation}
    \frac{1}{\omega-\hat K}
\end{equation}
shows that singularities of the Green's function in the
$\omega$ plane are determined by the spectrum of the BFKL kernel. In
particular, the rightmost singularity controls the asymptotic
high-energy behavior of the amplitude. Spectral properties of the kernel are consequently relevant to the
construction of the amplitude. 

\subsection{LO and NLO BFKL kernel eigenfunctions}
LO BFKL kernel eigenfunctions and eigenvalues are known and derived employing the the LO kernel conformal symmetry:
\begin{equation}
\varphi_{\nu,n}^{(0)}(\bm q)= \frac{1}{\pi\sqrt{2}}
\left(\bm q^2\right)^{i\nu-\frac12}
e^{in\phi}, \, \chi(\nu,n)=2\psi(1)
-\psi\left(\frac{n+1}{2}+i\nu\right)
-\psi\left(\frac{n+1}{2}-i\nu\right).
\end{equation}
Eigenfunctions form a complete basis in variables $\nu$ and $n$: $\nu$ controls the dependence on the transverse momentum
scale, whereas the integer $n$ is the conformal spin and determines the
azimuthal dependence.

The NLO kernel is no longer conformal; it is however possible to solve the eigenvalue problem perturbatively starting from the leading order, as shown by Chirilli and Kovchegov \cite{ChirilliKovchegov}:
\begin{equation}
\varphi_{\nu,n}(\bm q) = \varphi_{\nu,n}^{(0)}(\bm q)
\left[
1+\bar\alpha_s(\mu_R)
h_{\nu,n}(\bm q^2)
\right], \ \ \Delta(\nu,n)
=
\bar{\alpha}_s(\mu_R)\,\chi(\nu,n)
+
\bar{\alpha}_s^2(\mu_R)\,\chi_1(\nu,n) \;,
\label{eq:nloeigenfunctions}
\end{equation}
where
\begin{equation}
h_{\nu,n}(\bm q^2) = A_{\nu,n}\ln^2\frac{\bm q^2}{\mu_R^2}
+
B_{\nu,n}\ln\frac{\bm q^2}{\mu_R^2}.
\end{equation}
Then, eigenfunctions acquire logarithmic distortions in $\bm q^2$ associated to coefficients $A_{\nu,n}$ and $B_{\nu,n}$, which contain inverse powers of
$\chi'(\nu,n)$. Consequently, NLO eigenfunctions exhibit singular
behavior in $\nu$ and should be understood as distributions using the principal-value prescription.

Let us focus on how these eigenfunctions enter the amplitude computation. Going to NLLA introduces corrections in both kernel and impact factors, which should be included at NLO accuracy (for recent results see~\cite{Nefedov:2019mrg,Hentschinski:2020tbi,Celiberto:2022fgx,Fucilla:2024cpf,Nefedov:2024swu,Fucilla:2026ooq}). In the LO basis, the NLO kernel is
not diagonal and its action on the eigenstates contains 
non-diagonal contributions giving rise to derivative
operators with respect to $\nu$.

Restricting to the $n=0$ sector the NLLA amplitude in the LO eigenfunction basis can
be represented as
\begin{equation}
{\rm Im}\mathcal A_{\rm LO \ basis}
= \frac{s}{(2\pi)^2}
\int_{-\infty}^{+\infty}d\nu
\left(\frac{s}{s_0}\right)^{\bar\alpha_s(\mu_R)\chi(\nu)}
\alpha_s^2 c_1(\nu)c_2(\nu)
\Bigg[
1+\bar\alpha_s(\mu_R)\left(\frac{c_1^{(1)}(\nu)}{c_{1}(\nu)} + \frac{c_2^{(1)}(\nu)}{ c_{2}(\nu)} \right)
\notag
\end{equation}
\begin{equation}
+\bar\alpha_s^2(\mu_R)
\ln\left(\frac{s}{s_0}\right)
\left(
\bar{\chi}(\nu)
+
\frac{\beta_0}{8N_c}\chi(\nu)
\left[
-\chi(\nu)
+\frac{10}{3}
+i\,\frac{d}{d\nu}\ln\left(\frac{c_1(\nu)}{c_2(\nu)}\right)
+2\ln\mu_R^2
\right]
\right)
\Bigg],
\label{eq:lobasis}
\end{equation}
where $c_{1,2}$ ($c^{(1)}_{1,2}$) are the projections of the LO (NLO) components of impact factors onto the LO eigenbasis.

Alternatively, one can use the NLO eigenfunctions to construct a
spectral representation in which the NLO kernel is completely diagonal. NLO
impact factors are then projected onto the NLO eigenfunction basis following
\begin{equation}
\widetilde c_{1,2}(\nu) = c_{1,2}(\nu) + \mathcal{O}(\bar\alpha_s),
\end{equation}
and the NLLA amplitude takes the form
\begin{equation}
{\rm Im}\mathcal A_{\rm NLO\ basis}
=\frac{s}{(2\pi)^2}
\int_{-\infty}^{+\infty}d\nu
\left(\frac{s}{s_0}\right)^{
\bar\alpha_s\chi(\nu)
+\bar\alpha_s^2\chi_1(\nu)}
\alpha_s^2
\widetilde c_1(\nu)\widetilde c_2(\nu)
\left[1+\bar\alpha_s\left(\frac{c_1^{(1)}(\nu)}{\widetilde c_{1}(\nu)} + \frac{c_2^{(1)}(\nu)}{\widetilde c_{2}(\nu)} \right)\right].
\label{eq:nlobasis}
\end{equation}

A central result is that the two representations are equivalent within
NLLA accuracy, which is a verified property independent on the process \cite{Polizzi}. 
Nevertheless, both representations necessarily contain contributions
beyond NLLA, since they organize higher-order terms differently. The
two numerical implementations can therefore exhibit a different dependence on
the auxiliary scales entering the calculation, namely the energy scale $s_0$ and the renormalization scale $\mu_R$. This observation
provides the main phenomenological motivation for studying the NLO
basis.

\section{Phenomenology: NLO eigenfunctions applications}
\subsection{Electroproduction of light vector mesons}

As a first phenomenological application we consider the forward
electroproduction of two light vector mesons,
\begin{equation}
\gamma^*(p)+\gamma^*(p')
\rightarrow V(p_1)+V(p_2),
\qquad
V=\rho^0,\omega,\phi.
\end{equation}
We restrict ourselves to the dominant leading-twist transition from a
longitudinally-polarized virtual photon to a longitudinally-polarized
vector meson. The corresponding impact factor is known at NLO and can be written as
\begin{equation}
\Phi_{1,2}(\bm q)=\alpha_s D_{1,2}
\left[
C^{(0)}_{1,2}(\bm q^2)
+\bar\alpha_s C^{(1)}_{1,2}(\bm q^2)
\right],
\end{equation}
and, in the context of collinear factorization, it is given by the convolution of hard cross section for a collinear quark-anti quark pair production with the meson distribution amplitude. 

It is possible to test the stabilization effect of NLO eigenfunctions comparing NLLA amplitudes evaluated in the two basis and performing a scan in the external parameter space, searching for regions of reduced sensitivity by
means of the principle of minimum sensitivity (PMS) \cite{PMS}.

\begin{figure}
    \centering
    \includegraphics[width=0.5\linewidth]{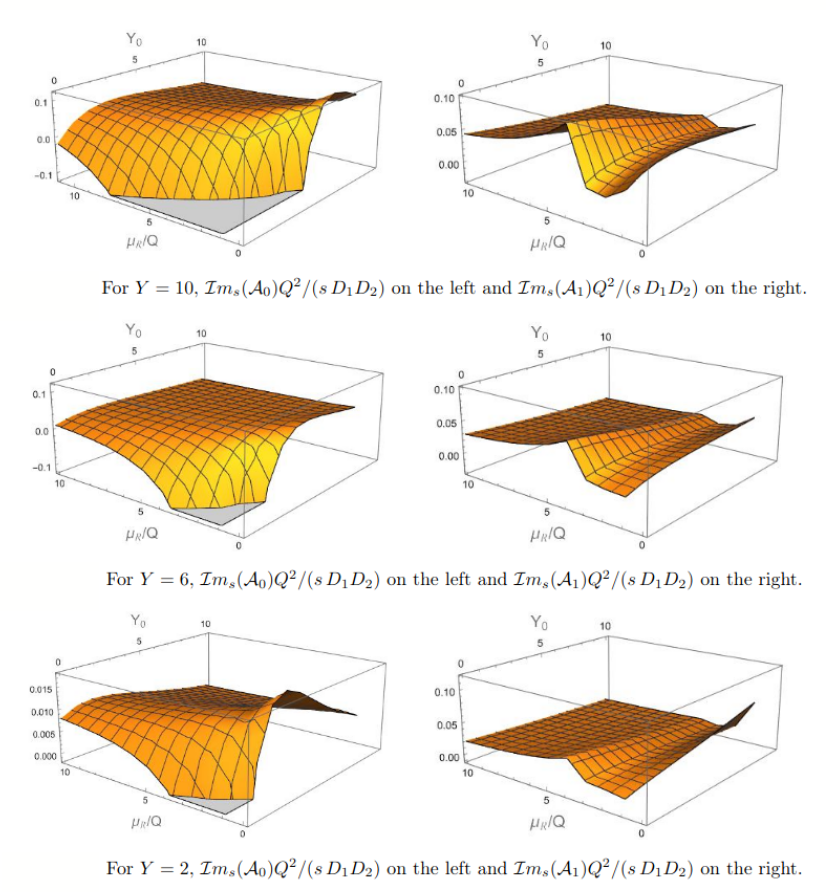}
    \caption{Scan on the external scales space in the standard representation.}
    \label{Standard}
\end{figure}

At $Q^2=24 \ \text{GeV}^2$ and at different rapidities, the LO-basis representation $\mathcal{A}_0$ was found to exhibit stability
only for values of $\mu_R$ and $s_0$ (or equivalently $Y_0=\text{ln}\bigg(\frac{s_0}{Q^2}\bigg)$) significantly larger than the
natural hard scale of the process \cite{Mesons}. The use of NLO eigenfunctions in $\mathcal{A}_1$
modifies the scale dependence and stationary points of the amplitude can
appear at smaller values of the renormalization scale, closer to the
physical scale set by the photon virtualities. This behavior has been tested for two amplitude representations, one called ``standard'' shown in Fig.\ref{Standard} and one which recasts the amplitudes into series, making explicit the resummation of leading and next-to-leading logs.

The vector-meson process
therefore provides a first indication that the NLO eigenbasis may offer
a more favorable organization of higher-order contributions in some
kinematic regions. Whether this behavior is universal or
process-dependent must, however, be investigated through further
applications.

\subsection{Mueller--Navelet jet production (preliminary results)}

Mueller--Navelet jet production \cite{MN1,MN2,MN3,MN4,MN5} provides a natural testing ground
for BFKL phenomenology:
\begin{equation}
p(p_1)+p(p_2)
\rightarrow \text{jet}(k_{J_1}) + \text{jet}(k_{J_2}) + \text{X}.
\end{equation}

In fact, the process is
characterized by the production of two jets separated by a large rapidity
interval and the large phase space available for additional semi-hard
radiation makes it particularly sensitive to BFKL dynamics.

At Born level jets are emitted back-to-back, while additional
radiation in the rapidity interval between them induces azimuthal
decorrelation. LLA BFKL overestimates these decorrelation and suffers from strong scale instabilities and consequently moving to NLLA is necessary to restore a reliable shape of the correlation spectra and improve stability. Expanding the differential cross section in
azimuthal coefficients $C_n$ is then useful in order to define azimuthal correlation:
\begin{equation}
\frac{d\sigma}
{dy_1\, dy_2\,d|\vec{k_1}|\, d{|\vec{k}_2|}\, d\phi_1\, d\phi_2}
=
\frac{1}{(2\pi)^2}
\left[
\mathcal{C}_0
+
\sum_{n=1}^{\infty}
2\cos(n\phi)\,\mathcal{C}_n
\right],
\qquad
\langle \text{cos}(n\phi)\rangle = \frac{\mathcal{C}_n}{\mathcal{C}_0}.
\end{equation}

In contrast to the vector meson case, several conformal spin components
contribute to Mueller--Navelet observables and NLO eigenfunctions should be studied beyond the $n=0$ sector. 

\begin{figure}
    \includegraphics[width=0.3\linewidth]{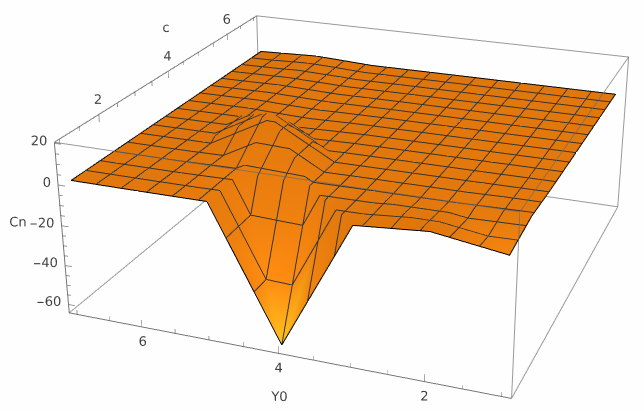}
    \includegraphics[width=0.3\linewidth]{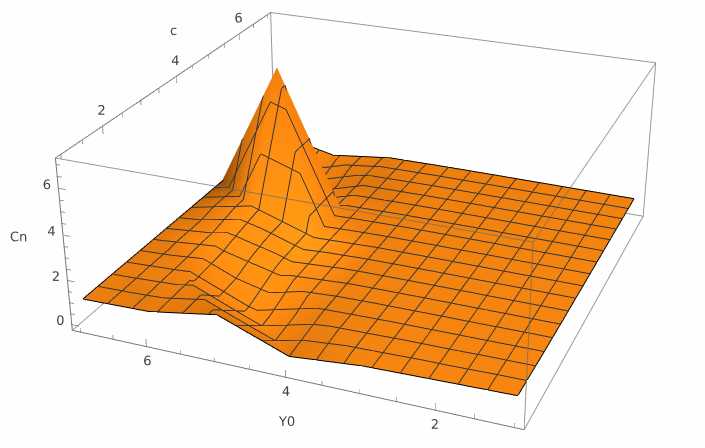}
    \includegraphics[width=0.3\linewidth]{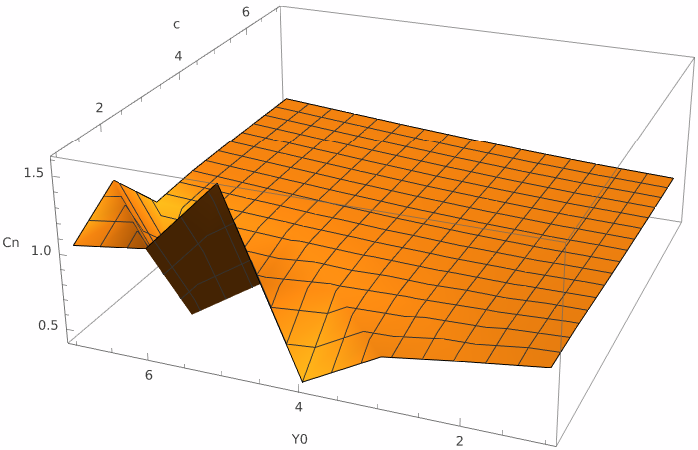}
    \caption{For $Y=6, \mu_R=\mu_F, \sqrt{s}=14 \,\text{TeV}, |\vec{k_{J_1}}|=|\vec{k_{J_2}}|=35 \, \text{GeV}, n=1$, plots of $\langle \text{cos}(n\phi)\rangle$ using the LO basis, the LO basis in exponential representation and the NLO basis respectively.}
    \label{MNjets}
\end{figure}

Preliminary results in Fig.\ref{MNjets} show
that predictions obtained in the NLO basis are more bounded under
parameter variations; however the best
stabilization is obtained with the exponential representation using the LO basis \cite{MN6}.

In Ref.\cite{Nefedov} a combination of BFKL NLO
basis, high-energy factorization and Sudakov resummation was used to set external scales without an optimization procedure (such as PMS or BLM). This indicates that for this process the combination of 
additional effects with respect to the simple use of NLO eigenfunctions may be required to obtain a significant phenomenological improvement. 

\section{Conclusions and outlook}

We have discussed the phenomenological use of NLO eigenfunctions of
the BFKL kernel. The main theoretical result is the process-independent
equivalence, within NLLA accuracy, between the representations of
high-energy amplitudes constructed using the LO and the NLO
eigenfunctions.

The two formulations differ only by terms beyond NLLA, which are
unavoidable in practical implementations and can affect the dependence
of observables on the renormalization and energy scales. Consequently,
the choice of basis, although formally irrelevant, can
have phenomenological consequences.

In the case of electroproduction of two light vector mesons, the NLO basis modifies
the amplitude stability region and may lead to stationary
behavior at scale values closer to the natural hard scale of the
process. On the other hand, preliminary studies of Mueller--Navelet jets show that, in this case, the improvement induced by the NLO basis is not evident without taking into account additional effects.

Further investigations are therefore required to establish under which
conditions NLO eigenfunctions improve perturbative stability. Future
directions include the application of collinear improvement to meson electroproduction, the combination of BFKL
evolution with high-energy factorization and Sudakov resummation for Muller-Navalet jets and applications to other relevant high-energy observables, such as forward-jet production in deep inelastic
scattering, di-hadron production at large rapidity separation and
heavy-quark production.

\acknowledgments

The work of A.Pa. and A.Po. is supported by the INFN/QFT@COLLIDERS project (Italy). The work of MF is supported by the ULAM fellowship program of NAWA
No.~BNI/\allowbreak ULM/\allowbreak 2024/\allowbreak 1/\allowbreak 00065
``Color glass condensate effective theory beyond the eikonal
approximation''.

\end{document}